\documentstyle[twocolumn,aps]{revtex}
\begin{document}
\draft

\title{Insulating and Conducting Phases of RbC$_{60}$}

\author{Michael C. Martin, Daniel Koller, Xiaoqun Du, Peter W. Stephens and
Laszlo Mihaly}
\address{Department of Physics, SUNY at Stony Brook, Stony Brook, NY
11794-3800}

\date{November 23, 1993; revised \today}
\maketitle

\begin{abstract}
Optical measurements were performed on thin films of Rb$_{x}$C$_{60}$,
identified by X-ray diffraction as mostly $x=1$ material.  The samples were
subjected to various heat treatments, including quenching and slow cooling
from 400K.  The dramatic increase in the transmission of the quenched
samples, and the relaxation towards the transmission observed in slow
cooled samples provides direct evidence for the existence of a metastable
insulating phase.  Slow cooling results in a phase transition between two
electrically conducting phases.
\end{abstract}
\pacs{PACS:  78.30.-j, 71.30.th, 63.20.-e, 78.66.Qh}
\narrowtext

The Rb$_{x}$C$_{60}$ system seems to have a very rich phase diagram. The
$x=0,3,4$, and $6$ phases \cite{structure} were identified shortly after the
discovery of superconductivity in intercalated C$_{60}$ \cite{discsuper}.
More recently a phase with one alkali atom per fullerene ($x=1$) was
discovered \cite{winter,poirier}.  With the exception of the superconducting
$x=3$ material, little is known about the electrical properties of the
compounds; in fact the exact composition of the nominally $x=3$ material
is also debated \cite{notx=3}.

The present work has been motivated by the conduction electron spin
resonance (CESR) experiments of J\`anossy {\it et al.} \cite{Janossy}
and by the NMR studies of Tycko {\it et al.} \cite{Tycko}.  J\`anossy
{\it et al.} studied a two-phase sample of Rb$_x$C$_{60}$, and decomposed
the CESR signal into two contributions.  They argued that the broader one
originates from the superconducting $x=3$ compound, while the narrower
one comes from another phase.  This signal exhibited a complex, history
dependent thermal behavior, with indications of a structural phase
transition in slowly cooled samples and a metastable insulating
quenched phase.  Earlier X-ray studies on the $x=1$ compound by Zhu
{\it et al.} \cite{Zhu} established the existence of a rocksalt structure
at high temperature, with a phase transition at around $120^o$C.
J\`anossy {\it et al.} tentatively assigned the narrower CESR signal to
the $x=1$ material, and argued that it is a metal at high temperatures.
On the other hand, Tycko {\it et al.} \cite{Tycko} interpreted NMR results
on RbC$_{60}$ in terms of a paramagnetic insulator.  We set out to explore
this compound by utilizing the direct link between IR transmission and
conductivity.

Optical measurements in the IR regime are helpful in the study of these
compounds in two ways.  First, the four IR active molecular vibrations are
sensitive to the electronic state of the molecule; the $F_{1u}(4)$ mode at
$1427 $cm$^{-1}$ in pure C$_{60}$ exhibits a characteristic change in
frequency and intensity as the doping proceeds \cite{Fu,Martin,Kuzmany}.
Second, mobile electrons in a metallic phase lead to an overall reduction of
the transmission $t$; in first approximation, $t\approx 1/\sigma _1^2$,
where $\sigma _{1}$ is the real part of the conductivity.  For frequencies
below the relaxation rate of the charge carriers, the IR transmission can
serve as a good indicator of the conductivity, although the accurate
evaluation of $\sigma (\omega )$ requires performing full
Kramers-Kr\"onig analysis of the data.

The two thin film samples used in this study were prepared in sealed
chambers, made to allow for controlled doping and for cooling and
heating over a wide temperature range \cite{Koller}.  The pristine
C$_{60}$ was deposited on high resistivity silicon substrates.
The thickness of the films were d=$2.5 \mu$m
for sample I and d$<1 \mu$m for sample II.
The doping was performed in a Bomem $MB-155$ FTIR spectrometer allowing
{\it in-situ} infrared transmission measurements at $2 $cm$^{-1}$
resolution during doping.  In order to achieve the best homogeneity
the films were kept at an elevated temperature $(125^o$C, to increase the
Rb diffusion rate) throughout
the process.  Rb was added in small quanta over several days, and the
dc resistivity and the infrared transmission spectra were monitored.
As demonstrated both experimentally \cite{Martin,Kuzmany} and theoretically
\cite{rice&choi}, the position of the
intramolecular $F_{1u}(4)$ vibrational mode is a good indicator of the
electronic composition of the doped fullerene film.  The appearance of
a single dominant resonance in the present sample (measured at $125^o$C,
see Figure \ref{fig1}, top panel) indicates homogeneity of composition;
the position of the peak, $1393$cm$^{-1}$, is clearly different from the
position of the $F_{1u}(4)$ peak in pristine C$_{60}$ $(1427$cm$^{-1})$.
The resistance of the sample dropped by several orders of magnitude, and
the transmission dropped by more than a factor of ten, indicating that
the doping produced an electrically conducting phase.

The doping method employed does not allow accurate determination of the
sample stoichiometry.  Therefore it is important to show that the conducting
phase seen in the IR transmission is not the $x=3$ material.  To this end
X-ray diffraction experiments were performed at the $X3B1$ beamline of the
National Synchrotron Light Source, at Brookhaven National Laboratory on the
very same specimens used in the IR studies.  The incident wavelength was
0.764\AA with the diffracted beam collimated by $0.1^o$ Soller Slits.
Spectra were taken at $125^o$C and room temperature in the angular range
covering the (220), (311) and (222) reflections of the cubic unit cell of
lattice spacing $\sim 14$\AA .  The small quantity of the sample and the
thick silicon windows limited the structural information available, but we
established unambiguously that the predominant phase in our specimens is
identical to the rocksalt phase seen by Zhu {\it et al.} \cite{Zhu}.  At
125$^{o}$C two peaks are observed at $2\theta =8.8^{o}$ and
$2\theta =10.3^{o}$.  These correspond to the (220) and (311) peaks,
respectively, of the rocksalt phase \cite{Zhu}.  Similar diffraction peaks
of the  $x=3$ material are at 8.6$^{o}$ and 10.1$^{o}$; these peaks were
very weak or totally absent in the two samples we studied.  After slow
cooling to room temperature the peak height decreased, and the width
broadened, indicating that the sample had  transformed in accordance
with the observations of Zhu {\it et al.} \cite{Zhu}.  Our results are
also in agreement with measurements on a slow cooled sample of
Rb$_1$C$_{60}$ in a more recent X-ray study by Chauvet {\it et. al.}
\cite{forro}.

In the IR experiments the samples were subjected to three thermal treatments
(A, B and C).  In (A) the specimen was thermalized at $225^o$C and it
was slowly cooled to room temperature at an average rate of
$5^o$C/hour, stopping at several fixed temperatures for taking the full
IR spectrum.  Treatment (B) was the inverse of (A), slowly heating from room
temperature to $225^o$C.  In treatment (C) the sample was quenched from
$225^o$C by pouring liquid N$_2$ on it.  In X-ray experiments performed on
powders of Rb$_1$C$_{60}$, the quenched state was seen to remain in the
rocksalt structure \cite{forro}.  After a few minutes of
thermalization at low temperature, the sample was quickly warmed to a preset
temperature below $60^o$C.  At each of these temperatures many IR spectra
were recorded, and the relaxation from the quenched metastable phase to
the equilibrium phase was studied.

The top panel of Figure \ref{fig1} shows spectra taken on sample I at
$125^o$C and at $-10^o$C during heat treatment (A).  The top panel
of Figure \ref{fig2} shows the temperature dependence of the transmission
for treatments (A) and (B) at a fixed wavenumber, chosen to be
$900 $cm$^{-1}$ to represent the broad band behavior, far away from
the sharp vibrational resonances.  Upon heating, treatment (B),
the specimen exhibits a phase transition around $120^o$C.  The
temperature hysteresis suggests that the transition is of first order.  As
confirmed by our X-ray data, this is the structural transition observed
by Zhu {\it et al.} \cite{Zhu}, although the transition temperature seems
to be sample dependent ($120^o$C \cite{Zhu,poirier}, $140^o$C \cite{Janossy},
$70^o$C \cite{Tycko}, and $90^o$C \cite{forro}).

The detailed evaluation of these transmission data indicate that the
material is an electrical conductor.  Reasonable plasma frequency and
relaxation rate values can be deduced in a Drude fit, but the two
parameters can not be determined independently.  For the high temperature
phase a resistivity of $\rho=18$m$\Omega $-cm was obtained, approximately
independent of the frequency in the measured range.  At low temperature
the conductivity is increased by about a factor of two.  Less than ideal
film quality can conceal the metallic character in the measured dc
resistivity \cite{Kochanski},
but the magnitude of the resistivity indicates that, although electrically
conducting, neither of the two phases is an ordinary clean metal.

Tycko et al. argues that in the high temperature phase the $^{87}$Rb NMR
data can be understood in terms of a paramagnetic insulator, with
electron spins localized to C$_{60}$ molecules \cite{Tycko}.  The dc
resistivity measurements on sample I seem to support this picture: it has
a high dc resistance, and decreases at higher temperatures.  However,
the dc resistivity can easily be dominated by grain boundaries.  (The
resistance seems to be very high even for the slow cooled state, for which
the NMR \cite{Tycko} and  CESR \cite{Janossy} data agree on the metallic
character.)  The IR transmission is less sensitive to grain boundaries.
Above 120C the transmission was found to be slightly decreasing (sample II)
as the temperature was raised.

In the slow cooled phase, close inspection of the $F_{1u}(4)$ line shows it
is split by about 16cm$^{-1}$ (1388cm$^{-1}$ and 1404cm$^{-1}$).  We believe
that the splitting is due to crystal field effects.  According to the NMR
results \cite{Tycko} the rapid rotation of the C$_{60}$ molecules
(characteristic of the high temperature phase) stops below the phase
transition.  In pristine C$_{60}$ the single $F_{1u}(4)$ line splits to
three lines (separated by $\sim 3$cm$^{-1}$) when the molecular rotation
ceases \cite{Narasimhan}.  The doublet seen in our measurement can be due to
stronger crystal fields with appropriate symmetry, related to the new low
temperature structure \cite{forro}, such that two of the three
$F_{1u}(4)$ modes are left
degenerate.  Another possible explanation is separation into
x=0 and x=3 phases.  However, the splitting we
observe is not consistent with the x=0 and x=3 IR $F_{1u}(4)$ peak positions
(1427 and 1363cm$^{-1}$ respectively).  Furthermore, Poirier {\it et al.}
\cite{poirier} have shown that Rb$_1$C$_{60}$
does not phase separate, even though K$_1$C$_{60}$ does \cite{poirierk}.

The lower panel of Figures \ref{fig1} and \ref{fig2} summarize
the temperature - and time-dependent responses in heat treatment (C).
Below 20C the quenched sample is an insulator, but it exhibits thermally
activated relaxation towards the stable (conducting) phase.  The
relaxation is illustrated in the lower panel of Figure \ref{fig1}, where
spectra taken on sample I
quenched to $-10^o$C are plotted.  The temperature dependence of the
transmission in the quenched state is better seen in the lower panel of
Figure \ref{fig2}.  Here the transmission of sample II at $900 $cm$^{-1}$
is plotted as a function of temperature.  The relaxation towards the
slow cooled state was recorded at several temperatures, repeating
treatment (C) each time.  The relaxation is well described by an
exponential time dependence with relaxation time $\tau$.  The inset in
the lower panel of Figure \ref{fig2} shows the relaxation time at
several temperatures.  For T$<-30^o$C, the relaxation is very slow and
the quenched phase is metastable.  Above the critical temperature of
T$_c=20^o$C the quenched state seems to be identical to the high
temperature conducting phase, and exhibits a (fast) relaxation towards
the slow cooled conducting state.

Immediately after quenching the transmission is high and vibrational modes
barely visible in the metallic state become pronounced.  The four symmetry
allowed IR active modes ($F_{1u}$) for the $x=1$ phase appear at 526, 575,
1183, and 1404cm$^{-1}$.  The resonances in Figure \ref{fig1}, lower panel,
at 1430, 1368 and $1341$cm$^{-1}$ correspond to the $F_{1u}(4)$ mode at
$x=0$, $x=3$ or 4, and $x=6$ respectively \cite{Martin,Kuzmany},
corresponding to a small amount of these phases in our film.  The
$F_{1u}(2)$ line at 566cm$^{-1}$ also originates from the $x=6$ phase.
These modes remain visible in both the high temperature and the slow
cooled metallic phases.  The $F_{1u}(3)$ mode around $1183$cm$^{-1}$
seems to be split, but this mode is not sensitive to doping
\cite{Martin,Kuzmany} or crystal field effects \cite{Narasimhan}.

In the quenched state, a number of new modes are visible in addition to
the four IR active vibrations allowed by the icosahedral symmetry of
the C$_{60}$ molecule.
Many of these modes can be assigned to Raman active $H_{g}$ and
$A_{g}$ modes seen in Rb doped C$_{60}$ \cite{modelist,Mitch}.  The
resonances around 480, 742, 838, 1080, 1192, 1243, and 1313cm$^{-1}$ are
near neutron scattering peaks \cite{neutron} and were identified as
fundamentals in a study on pristine C$_{60}$ single crystals \cite{xtal}.
In contrast to the resonances due to the
$x=3$ and $x=6$ impurity phases, all of these lines disappear as the
relaxation proceeds, suggesting that they belong to the metastable $x=1$
compound.  The Raman and other optically silent modes may acquire IR
activity if there is a symmetry breaking in the solid state as, for
example, in the neighborhood of impurities.  Other possibilities include
the dimerization seen in organic charge transfer salts \cite{Horowitz} or
the charged phonon effect observed in low dimensional solids \cite{Rice}.

For wavenumbers corresponding to energies higher than the gap in the
electronic spectrum, the IR transmission of an insulator is expected to
be similar to its metallic state transmission.  In our measurement the
transmission of the quenched insulating state stayed
high up to $4000$cm$^{-1}$, indicating a large band gap.  If the phase
transition is driven by an instability of the electronic system, the gap
should be related to the critical temperature by a factor on the order of
unity.  In conjunction with a transition temperature around $290$K,  the
gap in this metastable insulating material seems to be large,
$2\Delta > 8$k$_B$T$_c$.

In conclusion, we have established that the high temperature rocksalt
structure of the $x=1$ phase is electrically conducting.  Upon cooling, a
structural phase transition occurs leading to the splitting of the
$F_{1u}(4)$ intramolecular vibrational line.  The slow cooled material
is a better conductor.  When the sample is quenched to low temperature,
a metastable insulating state develops at temperatures below T$_c=290$K.
In this phase the gap in the electronic excitation spectrum is rather large,
for it does not show up in the measured frequency range extending up to
$4000$cm$^{-1}$ (320meV).

\acknowledgments

We are indebted to A. J\`anossy and L. Forro for valuable discussions and
for communicating unpublished CESR and X-ray data.  Discussions with
H. Kuzmany and D.M. Poirier are also appreciated.  This work has been
supported by NSF grant DMR9202528; the work at the NSLS is supported by
the Department of Energy Grant No. DEFG-0286-ER-45231.

\begin{figure}
\caption{The infrared transmission spectra of an RbC$_{60}$ film (sample I)
with various cooling histories.  Top panel shows the spectrum at $125^o$C
and slow cooled (treatment (B)) to $-10^o$C.  For the bottom panel, the
sample was quenched (treatment (C)) from $125^o$C to $-50^o$C.  The
temperature was then stabilized at $-10^o$C and spectra were obtained
as the sample slowly relaxed from the quenched state into its
equilibrium state.  The labels beside each curve indicate the time passed
since the quench.  For comparison, the slow cooled (treatment B) spectrum
is also reproduced here. The $590-650$cm$^{-1}$ region has a strongly
temperature dependent Si feature which is omitted.
Positions of the vibrational absorptions are indicated.}
\label{fig1}
\end{figure}

\begin{figure}
\caption{Infrared transmission at a fixed energy $(900 $cm$^{-1})$ for slow
cooling (treatment (A)), slow heating (B), and for the quenched state (C).
The upper panel shows the hysteresis between (A) and (B) for sample I.  The
lower panel presents the temperature dependence of the quenched state (C)
for sample II and its relation to the the slowly heated (A) data.
The inset shows the relaxation time, $\tau$ (plotted on a log scale in
minutes), at various temperatures.  The overall transmission difference
between the two samples is due to their different thickness.}
\label{fig2}
\end{figure}


\begin{references}
\bibitem{modelist}$H_g(3)$: 689, 705cm$^{-1}$; $H_g(4)$: 742, 761cm$^{-1}$;
$H_g(5)$:1082, 1093cm$^{-1}$; $H_g(6)$: 1243cm$^{-1}$; the $A_g(2)$ at
1451cm$^{-1}$ and $H_g(8)$ at 1527 and 1556cm$^{-1}$ are also observed.
Resonances at 504, 723, 816, 838, 1146, and 1192cm$^{-1}$ were not
identified with known fundamentals.
\end{references}
\end{document}